\documentstyle[12pt,aps]{revtex}
\begin{document}
\title{Ground state of the spin-$\frac{1}{2}$ Heisenberg antiferromagnet\\ 
on an Archimedean 4-6-12 lattice }
\author{P. Tomczak$^a$, J. Schulenburg$^b$, J. Richter$^b$, A.R. Ferchmin$^c$}
\address
{
$^a$Physics Department, Adam Mickiewicz University,\\
Umultowska 85, 61-614 Pozna\'{n}, Poland\\
$^b$Institut f\"ur Theoretische Physik, Otto-von-Guericke Universit\"at
Magdeburg,\\
P.O.B. 4120, 39016 Magdeburg, Germany\\
$^c$Institute of Molecular Physics, Polish Academy of Sciences,\\
Smoluchowskiego 17, 60-179 Pozna\'{n}, Poland\\
}
\maketitle

\begin{abstract}
An investigation of the N\'eel Long Range Order (NLRO) in the ground state of
antiferromagnetic Heisenberg spin system on the two-dimensional, uniform, 
bipartite lattice consisting of squares, hexagons and dodecagons is presented.
Basing on the analysis of the order parameter and the long-distance correlation
function the NLRO is shown to occur in this system.
Exact diagonalization and variational (Resonating Valence Bond) methods are applied.
\end{abstract}





Due to the recent renewal of interest in low dimensional quantum
antiferromagnetism, caused mainly by its possible connection with the
mechanism of the high-$T_c$ superconductivity, one has to notice a
great progress in the understanding of the nature of the ground state of
quantum Heisenberg antiferromagnets for low values of spin variables on
low-dimensional lattices. One of the basic issues in the
investigations concerning this subject is the question whether
a N\'eel Long Range Order (NLRO) exists in the ground state of an
antiferromagnetic spin-$\frac{1}{2}$ system on a given lattice and how it
can be destroyed.
This is also the question about the result of the nontrivial and subtle
interplay between quantum fluctuations and other mechanisms which can
destroy or stabilize NLRO in the ground state. At least two such mechanisms
seem to be relevant, namely the local singlet formation tendency and frustration. 
For example, the first mechanism which breaks the NLRO is present in the spin 
system on 1/5-depleted square lattice (being the prototype of the $CaV_4O_9$ lattice) 
and it manifests in continuous quantum
phase transition with critical exponents which seem to belong
to three-dimensional classical Heisenberg universality class\cite{{TKU},{TIU}}. 
On the other hand, in a case of a {\it generic}
model of frustrated antiferromagnet, see e.g., Ref. \cite{{Ri},{Ca}}, namely $J_1-J_2$ model, 
the growing frustration ($J_2/J_1$) gives rise to the continuous phase transition.
Remarkably, there also  may exist systems in which two above competing mechanisms
are being built in, like the spin system on Shastry-Sutherland lattice \cite{SS}
(being the prototype of $SrCu_2(BO_3)_2$ lattice). The question about 
the nature of the phase transition in this system remained for some time 
a puzzle and finally it came out that in the very small area of the parameter space
there exists a novel spin-gap phase between the dimerized and long range
ordered phases \cite{KK} and a continuous transition occurs in the vicinity 
of a discontinuous one.  Another example of this type is the $J-J'$ 
model (see, e.g., Ref. \cite{KR} and references therein).

Although it is rather widely accepted that 
spin systems with antiferromagnetic interactions on lattices with low
coordination numbers and frustrated ones 
are the best candidates for disordered ground state, the general question about
the NLRO remains not completely answered.

In this paper we focus on a spin-$\frac{1}{2}$ system with equal,
antiferromagnetic, nearest neighbor interactions:
\begin{equation}
H=\sum_{<i,j>}\roarrow{S}_{i}\cdot \roarrow{S}_{j}.  \label{eq:1}
\end{equation}
on one of the {\it Archimedean} lattices - on the
4-6-12 (square-hexagonal-dodecagonal - SHD) lattice.
Note, that in the spin system on the SHD lattice,
opposite to the honeycomb lattice (the same coordination
number), 
the nearest neighbors are not equivalent. This, in the natural way, 
favors the formation of the local singlets i.e., acts against the NLRO.
Since our earlier conclusion \cite{TR}
concerning the existence of NLRO was made basing on the results of 
a method which seem to overestimate the tendency towards NLRO, 
we present here a more extensive exact diagonalization and 
variational study.

To reexamine the problem of the existence of the magnetic order in the
ground state of the spin system on the {\it bipartite} SHD lattice
the RVB approach developed originally by Anderson \cite{And1} and
reformulated later by Liang, Doucot and Anderson \cite{{LDA},{Lia}} was applied.
This procedure seems to be well suited to spin systems on bipartite lattices.
At the beginning, however, let us describe the results of exact diagonalization
procedure applied to a 36-spin system with periodic boundary
conditions on SHD lattice, shown in Fig. 1. Those results were subsequently
used to estimate the quality of applying  the RVB method to the
same system. To diagonalize the 36-spin Hamiltonian the Lanczos algorithm
was applied. After using all possible point symmetries and spin reflection
the dimension of the $S_{tot}^{z}$ sector still 
amounted to  126,108,405. The ground state energy per bond of this
system is $E_{0}/bond=$-0.373118 and the correlation functions are collected
in Table I. In addition, in Fig. 2 the lowest energy levels of
this system vs. quantum numbers are presented. According to Anderson\cite{And2}
and Bernu et al. \cite{Be}  
the N\'{e}el long range order which breaks the rotational invariance in the
thermodynamic limit can occur if for small $S$ the lowest energy level for
each $S$ sector is linearly dependent on $S(S+1)$. This kind of dependence
is rather clearly seen in Fig. 2. This and the behavior of the averaged
correlation function with distance, seen in Fig. 3, form rather strong
evidence that the ground state in this spin system is long range ordered.
Let us also add that finite size analysis of the gap (based on ED results
for 12, 24 and 36-spin systems) gives small negative value (-0.055) of the spin
gap for infinite system and supports the above conclusion.

Now let us turn to the RVB method. It allows one to find a variational
ground state function for a given, finite spin system. Consequently, it is
possible to calculate, for a finite spin system, the expectation
values of the operators which, after the extrapolation to the thermodynamic
limit, can characterize the LRO in the ground state of an {\em infinite}
spin system. Let us remind the reader three essential steps of this method
as applied to quantum spin system on a bipartite lattice. Firstly,
the lattice is partitioned into two equivalent sublattices {\em A}
and {\em B}. Connecting all spins belonging to {\em A} sublattice with
arbitrary spins of the {\em B} sublattice and assuming that each pair of
connected spins is in a singlet state, i.e., \newline
$\mid \!i,j\,\rangle =\frac{1}{\sqrt{2}}(\mid \uparrow _{i}\downarrow
_{j}\rangle -\!\mid \uparrow _{j}\downarrow _{i}\rangle $, 
one produces a
covering $\mid \!c_{\alpha }\,\rangle =\prod_{i\in A,j\in B}\mid \!i,j\,\rangle$.
The system of all coverings forms, in fact, a new basis which is overcomplete
and not orthogonal: the amplitude of probability  
$\langle \,c_{1}\!\mid \!c_{2}\,\rangle$ 
that a system passes from $\mid
\!c_{2}\,\rangle $ to $\mid \!c_{1}\,\rangle $ is proportional to 
$2^{N(c_{1},c_{2})}$,
where $N(c_{1},c_{2})$ denotes the number of loops arising when
one draws the coverings $\langle \,c_{1}\!\mid $ and $\mid \!c_{2}\,\rangle $
simultaneously on the same lattice. Note that the Marshall sign 
rule is fulfilled automatically in this basis. Secondly, the ground state
variational function $\mid \!\Psi _{trial}\,\rangle $ is expanded into
 the  basis of the  functions $\mid \!c_{i}\,\rangle $ and
the positive coefficients (amplitudes) in this expansion are just the
variational parameters. At this point, however, two important assumptions
concerning amplitudes are made: the amplitude for a given covering has a
form of a product, i.e., factorizes with respect to singlets entering to
this covering.
An additional  assumption is that the singlets at the same distance
contribute to this product in the same way (form resonances - hence the name
of the method). Therefore, the trial wave function is assumed to be
\begin{equation}
\mid \!\!\Psi _{trial}\,\rangle =\sum_{\alpha }\prod_{i\in A,j\in
B}\,h_{ij}^{\alpha }\mid \!c_{\alpha }\,\rangle .  \label{eq:2}
\end{equation}
Finally, there follows a searching of the minimum of $\langle \Psi
_{trial}\!\mid \!H\!\mid \!\Psi _{trial}\rangle $ with respect to
variational parameters $h_{ij}^{\alpha }$ and the calculation
of the expectation values
of the desired operators in
the ground state of a spin system under consideration for those
$h_{ij}^{\alpha} $ which minimize $\langle \Psi
_{trial}\!\mid \!H\!\mid \!\Psi _{trial}\rangle $.
For small systems
this can be accomplished rigorously by taking into account the whole space
of coverings (e.g., for 12 spins there are 720 coverings, each covering
consisting of 64 Ising states), for larger ones 
by the  Monte-Carlo method, as proposed by Liang, Doucot and
Anderson \cite{{LDA},{Lia}}.

To make an optimal choice of the variational parameters $h_{ij}^{\alpha }$
we have calculated the variance of the ground state energy for small
clusters on some bipartite lattices. The whole basis of coverings was taken
into account. The best choice of $h_{ij}^{\alpha }$
which leads to a minimum  value of the variance in the ground state 
(with not too large dimension of the parameter space) 
is the following one: $(h_{AA}$, $h_{AB}$, $\sigma)$.
 Thus
$h_{ij}^{\alpha }=1$ for $r_{ij}=1$, $h_{ij}^{\alpha }=h_{AA}/r_{ij}^{\sigma
}$ for spins at the distance $r_{ij}$ belonging to the same sublattice, $%
h_{ij}^{\alpha }=h_{AB}/r_{ij}^{\sigma }$ otherwise, and $r_{ij}$ is the
Manhattan metric (the length of the  shortest path over bonds). All
the expectation values of operators were calculated for this choice of the
variational parameters. It seems to be important to choose not too high
a dimension of the variational parameter space. We have observed
that the minimum of $\langle \Psi _{trial}\!\mid \!H\!\mid \!\Psi
_{trial}\rangle $ is rather broad in the parameter space and small changes
of $h_{AA}$, $h_{AB}$, $\sigma $ lead to relative large changes of $m^{2}$.
It would mean that this method can 
also account for  some disordered singlet states slightly above the
ground state.

Table II presents the comparison between the exact diagonalization and
variational values of $E_{0}/bond$ and $m^{2}$ for 12 and 36 spin systems
with periodic boundary conditions.
 RVB method overestimates 
slightly the tendency towards LRO: variational values of $m^{2}$
are slightly higher - 0.07\% for 12-spin cluster and 5\% 
 for 36-spin cluster. The energy is reproduced very well: its
underestimation is only 0.4 \% for the 36-spin system. 
Those discrepancies result
from the singlet factorization assumption and their small values 
seem to indicate that it is a reasonable one. Let us also 
note that the parameters  $h_{ij}^{\alpha }$ 
decay much faster than the spin-spin correlations (for 36 spins 
$h_{AA}=0.950$, $h_{AB}=0.720$, $\sigma =1.54)$.
In Fig. 3 we also present the correlation functions vs. distance obtained
from the variational Huse-Elser ground state function \cite{{HuEl},{Car},{TR}}. 
They are 
overestimated in comparison to exact values  which forms
additional motivation to find the RVB ground state and to 
investigate the squared
magnetization calculated from the RVB ground state function in the 
thermodynamic limit.

Let us now describe our
results for larger systems. The variational values of $E_{0}$ and $m^{2}$ 
for 48-, 108-, and 192-spin systems with periodic boundary conditions
 are collected in Table II and 
their finite size analysis is presented in Figs. 4, 5.
 Since these quantities have 
a finite-size correction (for small $N$ corrections of higher orders may be
important), we decided to take into account only the data for $N=$48, 108 and
192 spins in the extrapolation.
The ground state energy per bond scales \cite{HN} like $N^{-3/2}$:
fitting the  data from Table II 
leads to  $E(N)=E_{\infty }+aN^{-3/2}$ with $E_{\infty }=-0.3688$
and $a=-0.8805$. Note that $E_{\infty }$ is slightly lower than that
obtained by Huse-Elser approach $(E_{\infty ,HE}=-0.3605)$ 
(see Ref.\onlinecite{TR}).  The square of order
parameter scales \cite{HN} like $N^{-1/2}$.
This leads to the following form 
of the square of sublattice magnetization as a function of 
$N$:
 $m^{2}(N)=m_{\infty }^{2}+bN^{-1/2}$ with $m_{\infty }^{2}=0.0648$ 
and $c=0.5136$. Note that $m_{\infty }$ is only
50\%  of its classical value (1/2) - which should be compared to
63\%  resulting from Huse-Elser ground state variational function.

Finally, in Fig. 6, the correlation function vs. the Euclidean distance is
plotted. It decays to about 0.09 for $r \sim 6$ and further almost does not change with
the distance. This
provides an  additional argument that the long-range magnetic order
persists in the ground state of this spin system.

To conclude, we have presented the results of the 
investigation of the ground state of the  
antiferromagnetic spin system on SHD lattice. The behavior of the low-energy
levels obtained from exact diagonalization, the value of $m_{\infty }^{2}$
and the finite value of the correlation function on higher distances 
represent  an evidence for the existence of two-sublattice N\'{e}el
long-range magnetic order in this system.

{\bf Acknowledgments} We acknowledge support from the Polish Committee for
Scientific Research (Project No. 2 PO3B 046 14) and from the Deutsche
Forschungsgemeinschaft (Projects No. 436 POL 17/9/00 and Ri 615/10-1). 
One of us (P.T.) thanks the Otto-von-Guericke University for support. 
Some of the calculations were performed at the Pozna\'n Supercomputer 
and Networking Center.

\begin{figure}
\label{pierw}
\caption{ The 36-spin system on bipartite square-hexagonal-dodecagonal lattice.}
\end{figure}

\begin{figure}
\label{drug}
\caption{ The lowest energy levels of the spin system from Fig. 1 vs.
quantum numbers $S(S+1)$. Straight line is the fit to the lowest energy in
each sector. }
\end{figure}

\begin{figure}
\label{trzec}
\caption{ The dependence of the sublattice correlation function 
on the Euclidean distance for the spin system shown in  Fig. 1. 
Comparison between exact diagonalization and variational results. }
\end{figure}

\begin{figure}
\label{czwart}
\caption{ Variational energy per bond $E_{0}/bond$ 
of  the spin system on the SHD lattice as a function of 
$N^{-3/2}$ extrapolated to the thermodynamic limit. Only three values (for $N$
= 48, 108 and 192 spins) were used in this extrapolation. }
\end{figure}

\begin{figure}
\label{piat}
\caption{ Squared sublattice magnetization $m^2 $ of  the spin system on 
the SHD lattice as a function of $N^{-1/2}$ extrapolated to the thermodynamic 
limit. Only three values (for $N$
= 48, 108 and 192 spins) were used in this extrapolation.} 
\end{figure}

\begin{figure}
\label{szost}
\caption{ The dependence of the sublattice correlation function 
on the Euclidean distance for the 192-spin system. }
\end{figure}

\begin{table}
\caption{The values of the correlation  
$\langle S^{z}_0S^{z}_i \rangle$ 
resulting from the exact diagonalization of
the 36-spin system depicted in Fig. 1}
\label{table1}
\begin{tabular}{cccccc}
$i$ & $\langle S^{z}_0S^{z}_i \rangle$ & $i$ & $\langle S^{z}_0S^{z}_i
\rangle$ & $i$ & $\langle S^{z}_0S^{z}_i \rangle$ \\ \hline
1 & -0.1381 & 13 & -0.0338 & 25 & -0.0360 \\
2 & 0.0561 & 14 & 0.0350 & 26 & 0.0386 \\
3 & -0.0533 & 15 & -0.0352 & 27 & -0.0436 \\
4 & 0.0561 & 16 & 0.0386 & 28 & 0.0350 \\
5 & -0.1033 & 17 & -0.0374 & 29 & -0.0354 \\
6 & 0.0521 & 18 & 0.0413 & 30 & 0.0356 \\
7 & -0.0474 & 19 & -0.0474 & 31 & -0.0354 \\
8 & 0.0365 & 20 & 0.0742 & 32 & 0.0335 \\
9 & -0.0354 & 21 & -0.1317 & 33 & -0.0354 \\
10 & 0.0356 & 22 & 0.0521 & 34 & 0.0413 \\
11 & -0.0389 & 23 & -0.0460 & 35 & -0.0460 \\
12 & 0.0340 & 24 & 0.0340 &  &
\end{tabular}
\end{table}

\begin{table}
\caption{The ground state energy per bond $E_{0}/bond$
and the squared sublattice magnetization $m^2$, for
some finite spin systems on SHD lattice. For the 12- and
36-spin systems the results of exact diagonalization are also included. In
the case of the 12-spin cluster the
variational values were obtained in the whole basis of coverings, for
larger clusters the Monte-Carlo method was applied.
Statistical errors, in parentheses, are the last two digits.}
\label{table2}
\begin{tabular}{cccc}
N &  & $E_0/bond$ & $m^2$ \\ \hline
12 & exact & -0.3850 & 0.2913 \\
& variational & -0.3850 & 0.2915 \\ \hline
36 & exact & -0.3731 & 0.1632 \\
& variational & -0.3718(15) & 0.1707(30) \\ \hline
48 &  & -0.3715(15) & 0.1402(37) \\ \hline
108 &  & -0.3698(16) & 0.1104(50) \\ \hline
192 &  & -0.3691(15) & 0.1044(54)
\end{tabular}
\end{table}

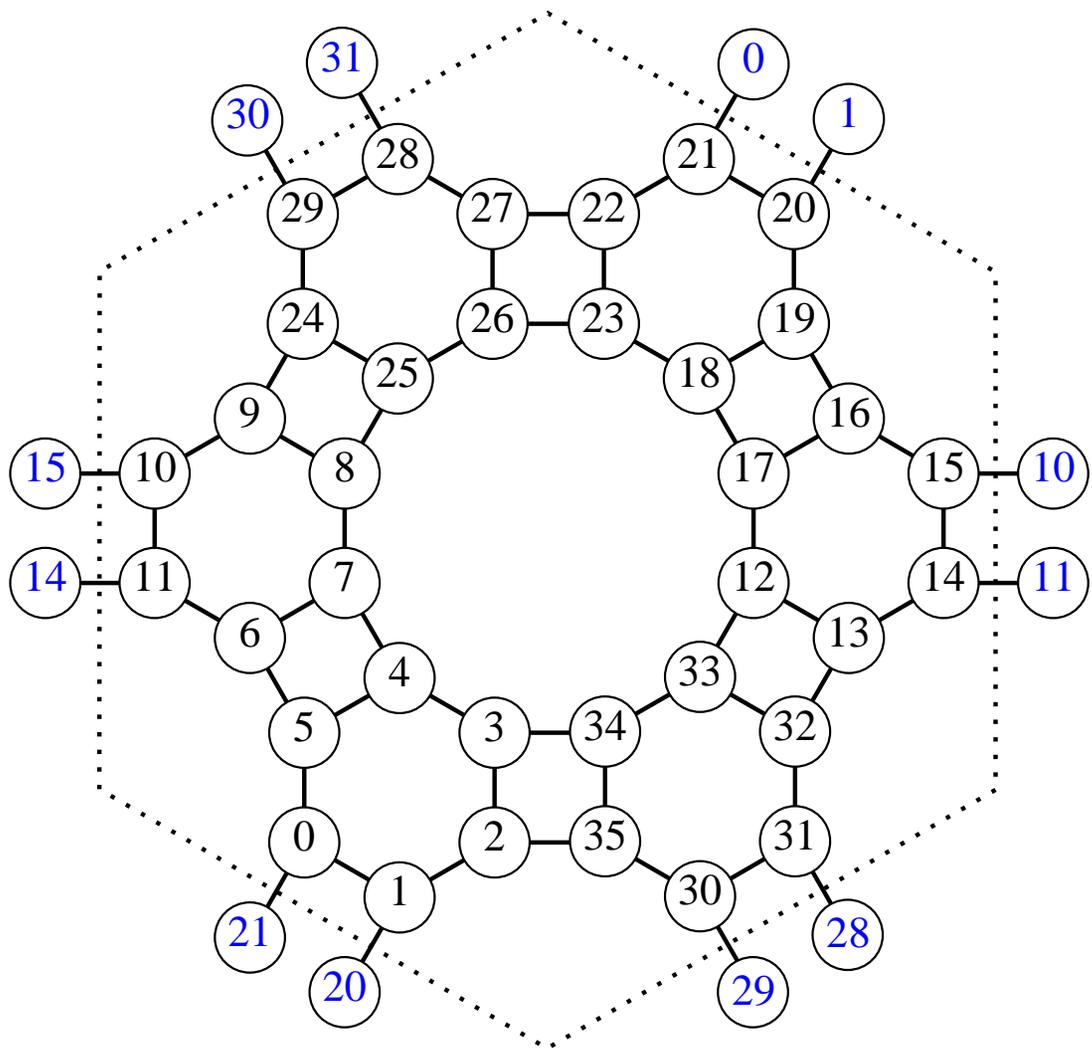
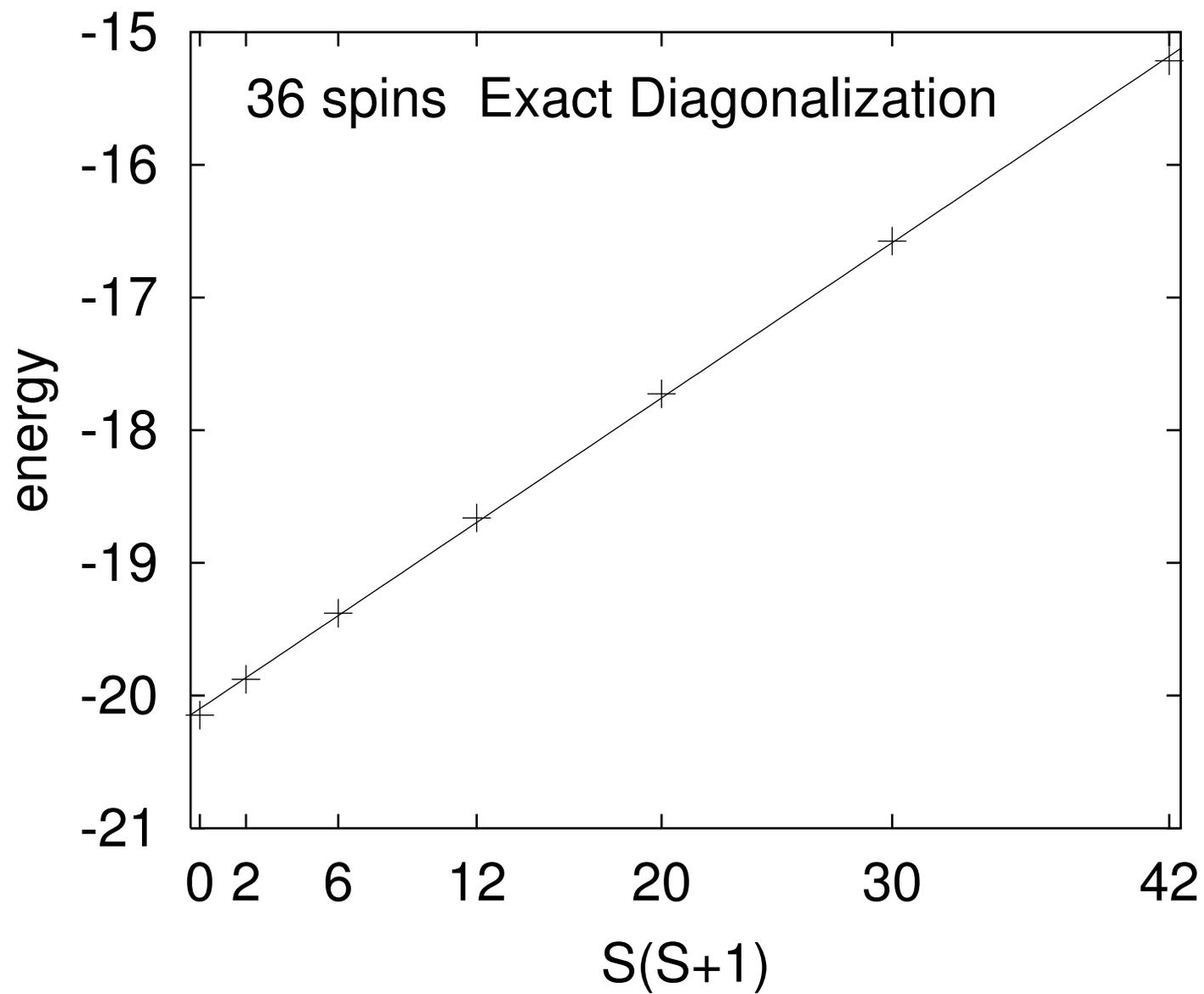
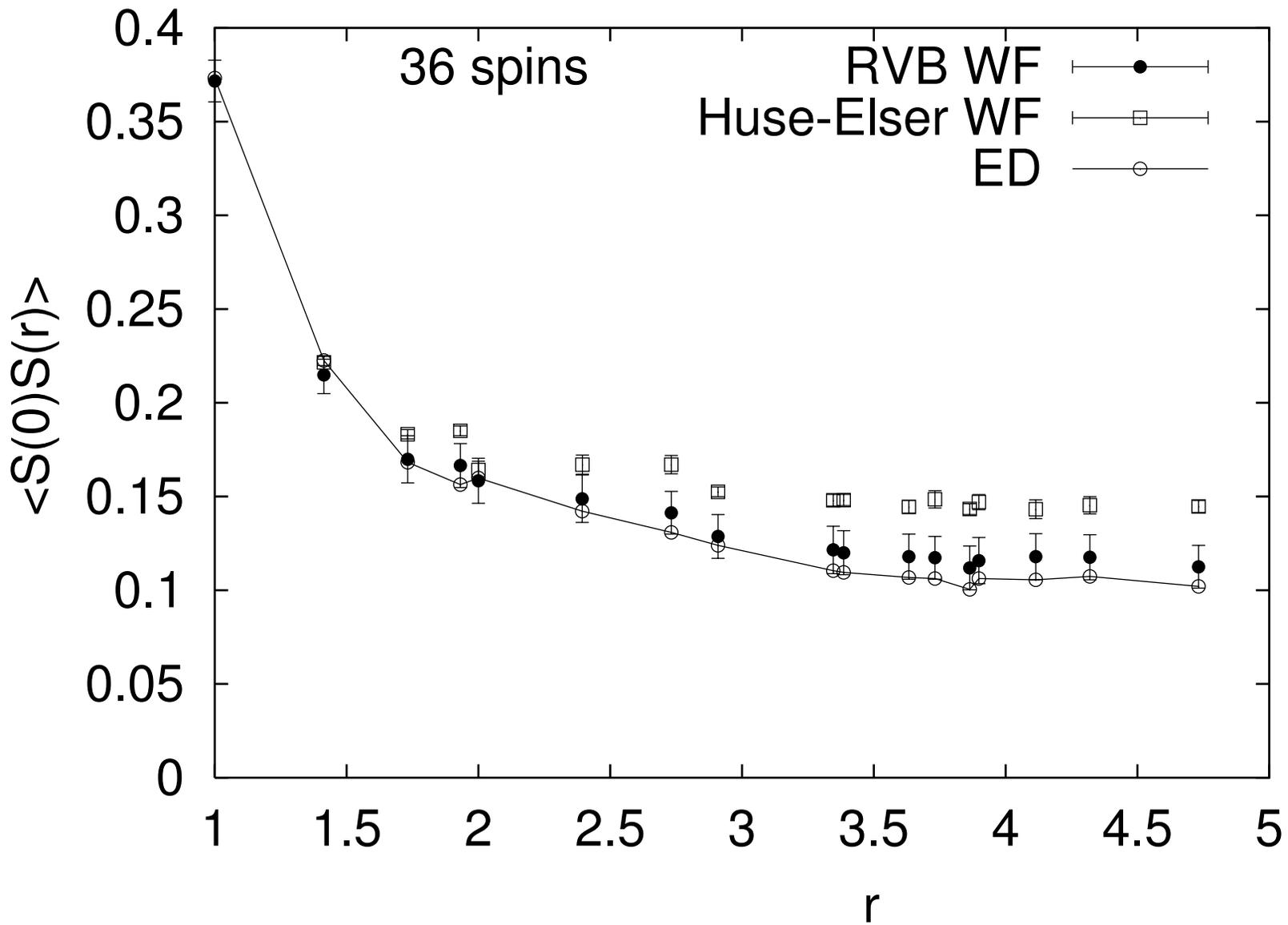
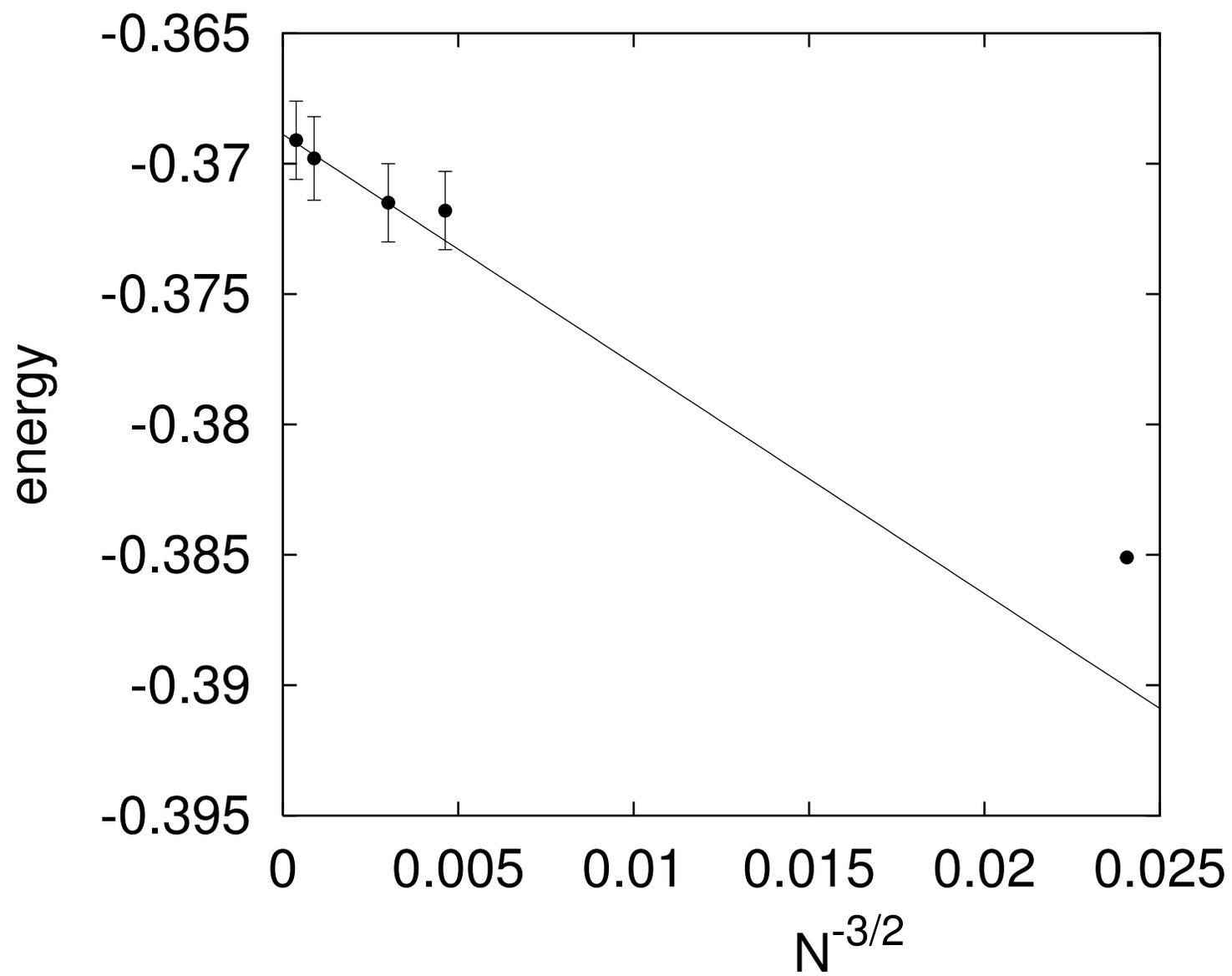
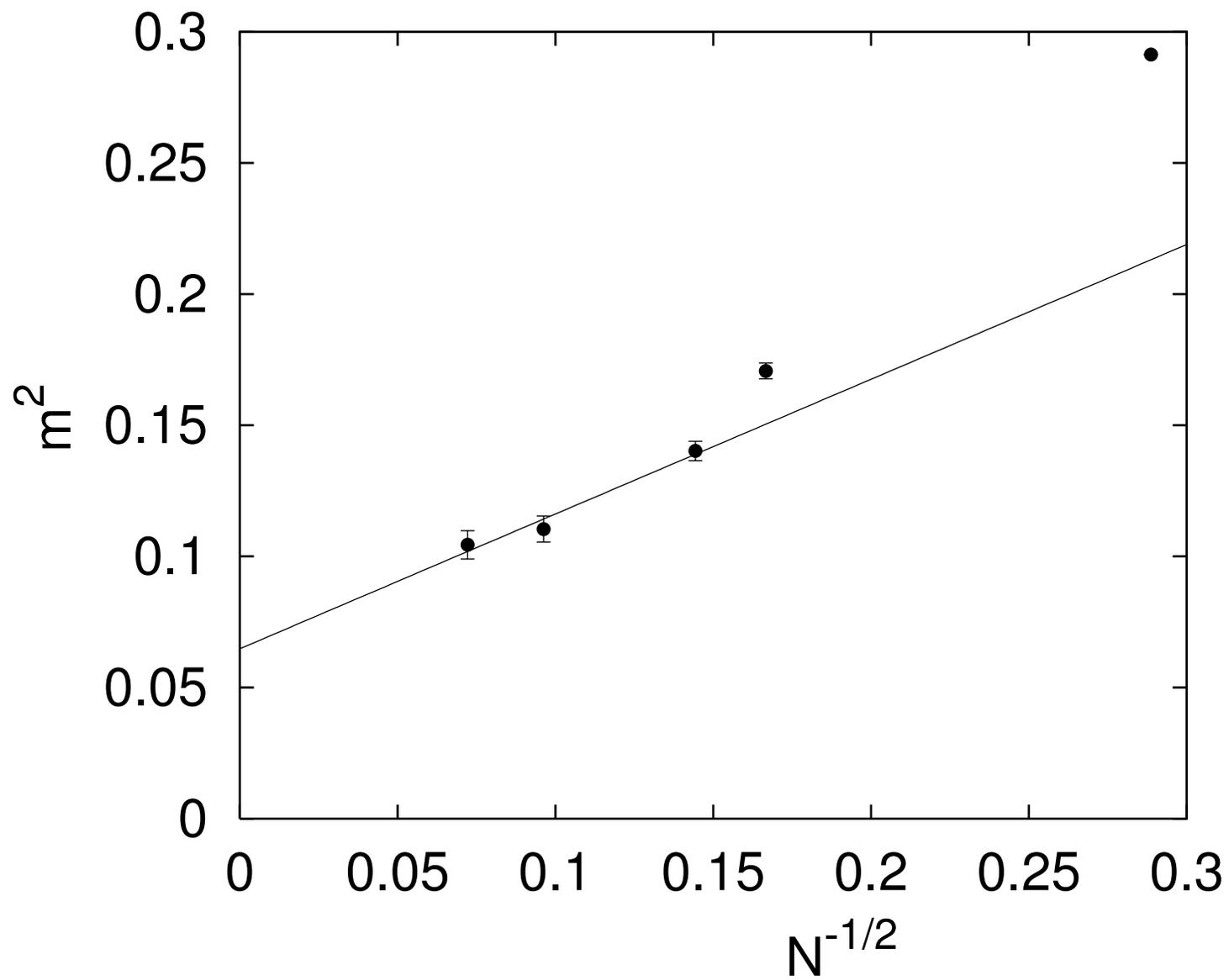
\end{document}